\documentclass[epj]{svjour}
\usepackage{epsfig,color,subfigure}
\newcommand{\bc}{\begin{center}}
\newcommand{\bareret}{d}
\newcommand{\beab}{\begin{Beqnarray}}
\newcommand{\ec}{\end{center}}
\newcommand{\eeab}{\end{Beqnarray}}
 \newcommand{\SMALLCAP}	[1]	{\caption[]{
 \begin{small} #1 \end{small}}}
 \newcommand{\vvec}	[1]	{\overrightarrow{#1}}
 \renewcommand{\vec}[1]{{\mathbf{#1}}}

 \newcommand{\bea}		{\begin{eqnarray}} 	
 \newcommand{\eea}		{\end{eqnarray}}
 \newcommand{\beann}		{\begin{eqnarray*}} 	
 \newcommand{\eeann}		{\end{eqnarray*}}

 \newcommand{\lrb}		{\left(}
 \newcommand{\rrb}		{\right)}

 \newcommand{\lab}		{\left\langle}
 \newcommand{\rab}		{\right\rangle}

\newcommand{\ii}{{\rm i}}
\newcommand{\isovol}{r}

\newcommand{\sep}{--}
\newcommand{\sepfrase}{---}
\newcommand{\horizon}{H}
\begin{document}
\title{Liquid markets and market liquids}
\subtitle{Collective and single{\sep}asset dynamics in financial markets}
\author{Gianaurelio Cuniberti
\thanks{e-mail: {\tt cunibert@mpipks-dresden.mpg.de}}%
\and Lorenzo Matassini 
} 
%
%
\institute{Max-Planck-Institut f\"ur Physik komplexer Systeme, N\"othnitzer
Stra{\ss}e 38, D{\sep}01187 Dresden 
}
\date{Received: / Revised version: }
%

\abstract{
We characterize the collective phenomena of a liquid market.
By interpreting the behavior of a no{\sep}arbitrage $N$ asset market in terms of a particle system scenario, (thermo)dynamical{\sep}like properties can be extracted from the asset kinetics. In this scheme the mechanisms of the particle interaction can be widely investigated.
We test the verisimilitude of our construction on 
two{\sep}decade stock market
daily data (DAX30) and show the result obtained for the {\em interaction potential} among asset pairs.
\PACS{
 {02.50.Sk}{Multivariate analysis} \and
 {89.90.+n}{Other areas of general interest to physicists}
 } 
} 
\maketitle
\label{intro}
Since the late 80s, with the introduction of electronic trading, huge quantities of financial data became available (or at least on sale) for both investment and research analysis. 
Quite unusually outside the natural science panorama, this novelty opened the way to test the reliability of theories and conjectures about the behavior of financial markets.
One of them, a paradigm for financial mathematics, is the random character of markets
\cite{Samuelson65}, 
that is unpredictability. 
It has recently been proved, nevertheless, that a certain degree of correlation is still present on extremely short time scales \cite{Lo91}.
Despite that, the intermediate scales are dominated by random behavior with L\'evy stable statistics of asset returns 
\cite{Mandelbrot63-and-Fama65}.
The possibility to extract information on the future evolution of a single asset by knowing a big enough ensemble of its past values matters indeed institutional traders, who can generally intervene on the market in real time (with delays smaller than few seconds). 
Their presence reduces at minimum time correlations and consequently
speculation possibilities.

Time dependence is however only one possible domain for surveying 
similar
patterns inside financial signals.
The other domain for correlation detection, whose exploration was greatly facilitated by modern computation facilities is the `spatial' one.
In fact, albeit much efforts are spent in studying correlations in the time dynamics of a {\em single} asset (see \cite{MS99} and \cite{BDLeBS96} for a digest of the recent physicist and economist approach, respectively), there are many applicative and fundamental reasons for understanding deeply spatial, commonly referred as {\em multivariate}, correlations.
A financial market is not simply a juxtaposition of different prices which are organised on an independent basis, but rather a complex system of interacting constituents 
\cite{Fama98}. 
The latter are then monitored by sampling single prices with respect to an arbitrary currency. Hence the study of correlations among different asset time signals is of peculiar importance. By the way, this is also the case in many problems involved in the modern risk management theory, where the composition of a certain portfolio strongly depends on the movements of different underlying assets. 
On a more fundamental level, the interesting issue is the comprehension of how price changes can be separated, with a sufficient degree of confidence, in {\em single asset}{\sep} and {\em collective}{\sep} behavior.

Since the Markowitz's work on the theory of optimal
portfolio \cite{Markowitz59}, much effort has been spent to characterize correlation matrices of financial assets \cite{EG95}. 
In recent contributions, different physics concepts have been adopted to endeavor this type of problem, mainly because the study of correlations represents a paradigm of a wide class of physical problems for which powerful tools have been developed.
A bivariate analysis of the futures on the German and Italian bonds showed
that despite the perfect uncorrelation of the single tracks, the
crosscorrelation of the two signals was significantly non zero: the signals considered described two random, but similar, processes \cite{CRS99}.
This behavior emerges quite generally in the stock market,
where certain asset clusters `move' in a particularly correlated way with respect to remaining titles. 
Using equal time cross{\sep}correlation matrices and several
physics{\sep}borrowed
tools such as the random matrix theory, these conjectures have been
quantified \cite{LCBP99-and-PGRAS99-and-DGRS99}.
In a recent study, the structure of a $N$ stock market has been investigated as regarding the multivariate structure in a global window period \cite{Mantegna99}.

In this paper, we propose a method to investigate asset correlations by interpreting asset growth rates as observables of a particle system scenario. This idea is carried out by introducing a formal map between the logarithmic returns and the distances among gas particles.
The strength of this analogy resides in the possibility to separate collective
motion from the single asset dynamics through the investigation of mutual interactions among titles.
Wielded by the theory of liquids, we can study the thermodynamics of the
system and interprete its temperature as a measure of spatial volatility, as compared with the more familiar (temporal) volatility. The 2{\sep}asset interacting potential is then calculated on the isothermal (isovolatile) market.
In the remainder of this paper a time dependent asset{\sep}distance and a moving frame model are introduces. The implementation of this scheme is performed on daily stock 
market data taken among the 30 most capitalized titles forming the {Deutscher Aktien} indeX (DAX30) in the 
period 30 Dec 1987 to 7 Mar 1995 (1800 trading days). 
To maintain a continuity of quotation, we have selected the maximal subset of 23 assets which, in the above mentioned period, remained in the DAX30 basket and did not split.
Our discussion and comments conclude the paper.

\label{sect:assdist}
As a general starting point, we consider a collection of asset, which is a suitable subpart of titles in a stock market (better if one representative for every economic sector), a collection of currency prices, or any combination of them.
The value of the {\em asset} $\Omega_i$ at time $t$, is expressed in unity of asset $\Omega_j$ by means of conversion factors $P_{ij}(t)$:
\bea
 \Omega_i (t) = P_{ij} (t) \Omega_j (t).
 \label{eq:convfact}
\eea
The indices $i$ and $j$ span all $N$ considered assets
forming the {\em market}. 
By writing eq.~(\ref{eq:convfact}) for another couple of indices, a no{\sep}arbitrage equation for a liquid market is obtained
$P_{ij} = P_{ik} P_{kj}$.
Its multiplicative symmetry 
is reflected in a corresponding additive symmetry of the logarithmic returns
\bea
\bareret^\alpha_{ij} (t) = \frac 1 {\tau_\alpha} \log \lrb \frac{P_{ij}(t)}{P_{ij} \lrb t-\tau_\alpha \rrb} \rrb, 
\label{dist}
\eea
where $\tau_{\alpha\le \horizon}$ is a collection of $\horizon$ time horizons.
 The rescaling of the log{\sep}returns to the considered time horizon is solicited by its interpretation;
in the idealized limit of prices with (deterministic) growth laws, we get 
$P_{ij} (t) \propto \exp \lrb \bareret_{ij} t \rrb $,
so that the quantity defined in eq.~(\ref{dist}) turns out to be the growth rate between asset $i$ and $j$,
independently on the time horizon.
The latter can be considered as a long term limit when one refers {\sepfrase}for example{\sepfrase} to prices of stocks with respect to currencies.
In the opposite limit of extremely small returns (which eventually corresponds to short time lags), $\bareret^\alpha$ is the rate of the absolute return,
$\bareret = \Delta P / ( P \Delta t )$, obtained by logarithmic expansion.

As position~(\ref{dist}) points out, the display of the time series $P_{ij}$ by
arranging them in the $\horizon$ dimensional variable $\vec{\bareret}_{ij}$, gives a
natural {\em embedding} for a dynamical system oriented analysis \cite{SYC91}. 
This is not difficult to understand when thinking that the log{\sep}return on a
certain time horizon $\tau^*$ is proportional to the average of log{\sep}returns on
sub{\sep}multiples of $\tau^*$. 
Thus the component $\bareret^{\alpha^*}$ can be written as a linear
combination of delayed components $\bareret^{\alpha < \alpha^* }$.
The no--arbitrage symmetry of the log{\sep}returns  hints for the
further identification of $\vec{\bareret}_{ij}$ as an (oriented) distance vector between asset $i$
and $j$; in fact ({\em a.}) $\vec{\bareret}_{ii} \equiv \vec{0}$, ({\em b.}) $\vec{\bareret}_{ij} = - \vec{\bareret}_{ji}$, ({\em c.}) $\vec{\bareret}_{ij} = \vec{\bareret}_{ik} + \vec{\bareret}_{kj}$.
%
%
\begin{figure}[t]
\begin{center}
\epsfig{file=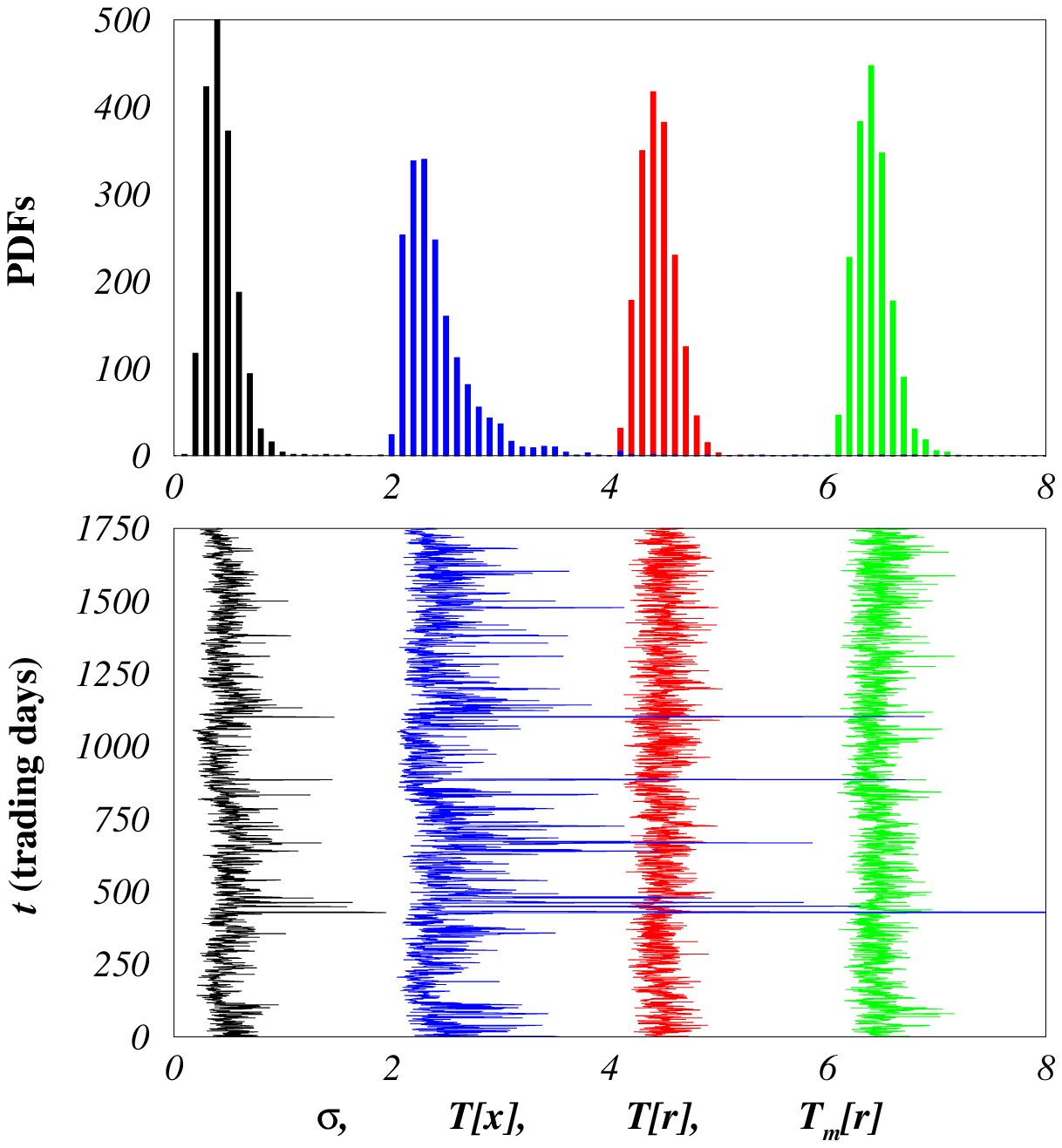, width=0.99\columnwidth}
\end{center}
\SMALLCAP{\label{fig:temp}Time dependence of the correlated volatility $\sigma$, and the temperatures $T$ (shifted as a visual aid) relative to the $x$\sep \ and $r$\sep coordinates (a); the corresponding PDFs are in panel b. 
All the calculations refer to four horizons ($\horizon=4$) of 1, 5, 20, and 250 market days.}
\end{figure}
It is easy to see that any norm in a $\horizon$ dimensional euclidean space (in the present work we choose the canonical one)
induces a well defined distance\footnote{Here the three distance{\sep}defining axioms are obtained simply by properties {\em a}.{\sep}{\em c}.} $\Vert \vec{\bareret}_{ij} \Vert$ between asset $i$ and asset $j$. 
As an intrinsic character of financial markets no asset can be regarded {a priori} as an absolute quantity, that is why we ended up only with mutual distances among asset. Nevertheless some truly single asset property can be extracted by the 
symmetry of the problem and interpreted consequently.
The matrix $\vec{\bareret}_{ij}$ is skew symmetric, ergo diagonalizable; its 
spectrum is entirely on the imaginary axe \cite{Gantmacher90}. 
One of its three different eigenvalues is zero and the corresponding eigenspace is
orthogonal to $\vvec{\vec{1}} = (\vec{1},\vec{1}, \dots, \vec{1})^{\rm t}$ and
$\vvec{\vec{x}} = (\vec{x}_1,\vec{x}_2, \dots, \vec{x}_N)^{\rm t}$, where
\bea
\vec{x}_i \equiv \frac 1 N \sum_{j=1}^N \vec{\bareret}_{ij}; \label{eq:posit} 
\eea
here the arrows indicate super{\sep}vectors (vectors in a $\horizon \times N$ space).
The two remaining eigenvalues are $\pm \ii N \sigma$ corresponding to the
eigenvectors $\vvec{\vec{1}} \mp \ii \vvec{\vec{x}} / \sigma$, where
\bea
\sigma \equiv \frac 1 N \sqrt{\sum_{1\le i<j \le N} \left\Vert {\vec{\bareret}_{ij}}\right\Vert^2}.
\label{eq:standdev}
\eea
%
%
\begin{figure}[t]
\begin{center}
\subfigure{\epsfig{file=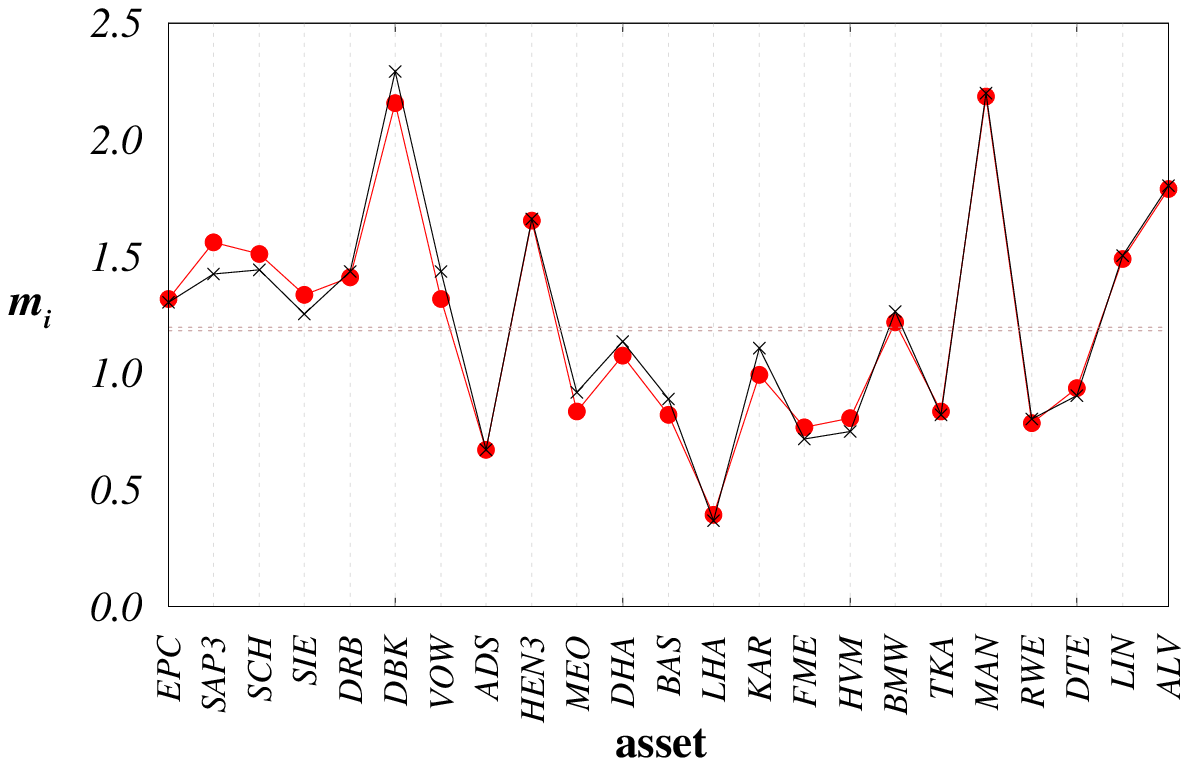, width=0.99\columnwidth}}
\end{center}
\SMALLCAP{\label{fig:masse}Masses $m_i$ as calculated after eq.~(\ref{eq:masses}) versus the asset label \cite{DB}. Crosses indicate $x${\sep}frame calculations with mean $1.195$, circles the $\isovol${\sep}frames with mean $1.184$. 
}
\end{figure}
In particular let us observe that $\vec{x}_i - \vec{x}_j = \vec{\bareret}_{ij}$,
to say that we have introduced a frame in which every single asset is assigned
to an absolute position:
the problem of the behavior of the $N$ assets of the market is now translated
to a physical problem of $N$ interacting particles (a liquid) in $\horizon$ dimensions, with coordinates $\vec{x}_1,\vec{x}_2, \dots, \vec{x}_N$.
At time $t$, $\vec{x}_i (t)$ is the $\horizon$ dimensional position of particle $i$.
Note that, according to its definition, the distance
between two assets is zero when the price of one with respect to the other
remains constant.
Furthermore, it is easy to check that the $\vec{x}$'s vectors are centered, hence the positions $\vec{x}_i$ are referred to a coordinate frame which attributes to the center of mass of our liquid a trivial dynamics. 
From the financial point of view, it states the {\em closure} of our system: the $N$ assets are watched as complementary, with zero overall return.
This does not mean that the applicability of the present construction is restricted to those market where
this property is nearly fulfilled (as an example in the foreign exchange).
In stock markets, which experience escape and retention events that is  
positive and negative return periods, the $\vec{x}$ are automatically
selected within a neutral frame which keeps track of the particle
cloud. 
Of course nothing prevents from 
starting the analysis of an extended market with a huge number of constituent assets.
Some of them would follow similar dynamics by evolving in a closer cluster
with respect to others. This could help in order to reduce
$N$ to a lower number without losing the basic features of the liquid behavior
\cite{MC00}.

Coming back to the map construction it is easy to show that 
as a consequence of the centered character of the $\vec{x}$'s coordinates,
$\sigma$ is exactly their standard deviation.
Its $\horizon$th power is a measure of the {\em volume} of our system. 
The financial counterpart of it is what we call {\em correlated volatility},
so to stress that it is a quantity merely connected to the spatial
interactions of the particles at a certain time. 
As the usual volatility takes into account the 
{\em temporal} variability of an analyzed fixed asset, we are here referring to a
measure of a {\em spatial} variability of a group of interacting assets at a fixed time.
Moreover, even after the compensation of the split discontinuities, the 
correlated volatility shows clusterization around bubble and crash periods \cite{MC00}. 

We take now advantage of
the structure of the eigenvectors of the distance matrix and rescale the ${x}$ coordinates to volume renormalized ones $\vec{\isovol}_i \equiv \vec{x}_i / \sigma$, their difference is accordingly $\vec{\isovol}_i - \vec{\isovol}_j = \vec{\bareret}_{ij} / \sigma$, so that the two non trivial eigenvectors are indeed $\vvec{\vec{1}} \mp \ii \vvec{\vec{\isovol}}$.
The $\isovol${\sep}frame, being the solution of the eigenvalue problem for the distance matrix, is a volume preserving frame. 
Once the volume of the system is stabilized, one may wonder which is the
dependence of the liquid temperature on time.
Thus, by analyzing the empirical behavior of the ensambled averaged square (finite difference) velocities 
${\vec{v}}_i (t)= ({\vec{\isovol}_i (t) - \vec{\isovol}_i (t-\tau_1)} ) /{\tau_1}$,
we found that the
${\isovol}${\sep}system
is thermostated at a fix temperature 
$T = \lab \lab {v}^2_i (t)\rab_i \rab_t / H  $ ;
the {\em correlated volatility} is therefore a measure of the {\em temperature} of our system.
%
%
\begin{figure}[t]
\begin{center}
\subfigure{\epsfig{file=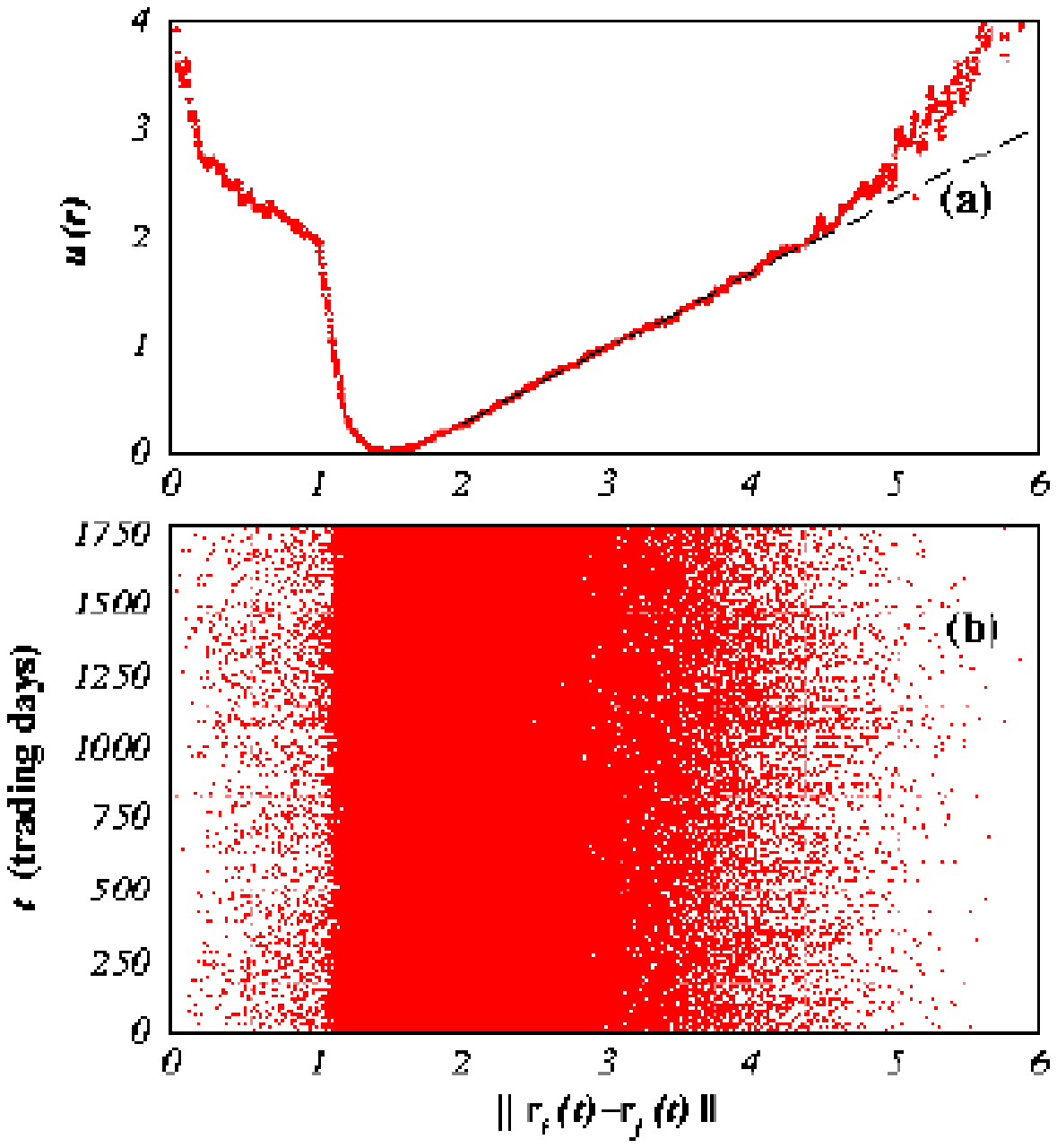, width=0.99\columnwidth}}
\end{center}
\SMALLCAP{\label{fig:pot} (a) Plot of the pair potential $u(r)$ for the whole data set over four horizons ($\horizon=4$) of 1, 5, 20, and 250 market days, and (b) the time distribution of the inter{\sep}asset distances (a better resolution figure is available upon request).}
\end{figure}
Fig.~\ref{fig:temp} shows this fact: in panel (b), it is plotted the time dependence of the correlated volatility $\sigma$ and of the temperatures $T[x]$ and $T[r]$ calculated by averaging the square velocities in the $x${\sep} and $\isovol${\sep}frame, respectively. 
In order to contrast the results, the time averages of $\sigma$ and $T[x]$
are rescaled to T (the time average of $T[r]$). 
The scale of $T$ is in fact fixed by the underlying assumption of an unitary
Boltzmann's constant.
To check possible ergodicity properties of the system we have also
analyzed the time averaged square velocities of the single assets and
extracted from them mass terms
\bea
m_i = \frac {\horizon T} { \lab {v}^2_i (t) \rab_t} .
\label{eq:masses}
\eea
Fig.~\ref{fig:masse} shows that the masses are only slightly affected by the reference frame used to calculate them. This indicates that they are an intrinsic property of the asset regardless of the kinetics details. To prove this statement, we have plotted, in Fig.~\ref{fig:temp}, the correction to the temperature due to the asset masses
${T}_m[r] =  \lab m_i {{v}^2_i (t)}\rab_i /H$.
In order to investigate the nature of the interaction of the particle system
under study, we have calculated the two point correlation function \cite{Martynov92} 
\beann
g (r) = \frac 2 {N (N-1)} \sum_{i<j} \lab 
\delta \lrb r- \left\Vert \vec{r}_{i} (t) - \vec{r}_{j} (t)\right\Vert \rrb 
\rab_t , 
\eeann
and the related pair potential $u (r) \propto - \log {g (r)}$.
In Fig.~\ref{fig:pot}., the potential $u(r)$ is shown.
The great distance tail of $u(r)$ is linear (correlation coefficient$=0.9994$,
for a regression in the region $2 < r< 4$ over 446 points giving the line $u =
a r + b$, with $a=0.689 \pm 0.001$ and $b= -1.101 \pm 0.004$)
indicating the strong long range attraction of the market liquid. On the other
hand at small distances two different behaviors emerge. By decreasing the
asset{\sep}asset distance an equilibrium point is reached. At smaller distances
a barrier is presented, followed by a region corresponding to less
intense repulsive forces. We interprete it as a signature of the inhomogeneity
of the system, which allow at small distances the formation of privileged
pairs (clusters). As a consequence, we expect that in a wider market (here we consider the
quite diversified but small pool of the DAX30 assets) this tendency could
even be more pronounced. 


To conclude,
we have introduced an interpretation scheme for the returns of a $N$ asset market. Here, the validity of a no{\sep}arbitrage condition is guaranteed by the assumed liquid character of the market. By implementing the embedding, naturally prompted by the structure of the returns, we have been able to map the financial signals in positions of particles of an interacting gas (a liquid). Therefore by the only means of the geometrical construction, we have given purport to the eigenvalues and eigenvectors of the distance matrix diagonalization problem as the correlated volatility and thermostated coordinates respectively. 
One of the strength points of this method is its easy generalizability
to the case of great $N$, albeit here we have restricted our
analysis to a relative small asset market.

On the other hand a word of caution is needed in a great $N$ market. The results presented here share the plain assumptions of isotropy and homogeneity of the market liquid. Indeed they should become weaker for very large and differentiated markets. There, the pair potential introduced here is supposed to maintain the same great--distance properties (linearity).
At the low--distances (where clustering emerges), in analogy to what is done in the study of ionic liquids \cite{McDonald91}, a generalized pair potential could be introduced in order to include both anisotropy, cluster formation, and specie diversification: these issues are under investigation and will be published elsewhere~\cite{MC00}.
Besides, this approach is straightforwardly employable for {\em time dependent clustering}. A procedure similar to the one adopted to organize static distances between assets in hierarchy trees, given in Ref.~\cite{Mantegna99}, could be generalized to the time dependent distance matrix~(\ref{dist}).

From the financial perspective, the construction presented in this paper is following a sort of {\AE}sop's {\em the fox and the grapes} strategy. It is easy to despise what one cannot get and in quantitative finance the scarce goods are the rare events. Since there is no methodology to deal with misprediction given by the insufficiency of statistics, we try to wash it out from the dynamics by exploiting the symmetries of the problem.  After all, as we have shown, the calm (non-bubble, non-crash) side of financial markets has anyhow a lot to say.

\label{sect:ack}
The data for the empirical analysis were kindly provided by Deutsche B{\"o}rse AG.
We would like to acknowledge fruitful discussions with A. Amici, F. Lillo, R. Mantegna, E. Scalas, and U. Tartaglino.

{

}


\end{document}